\documentclass[prb, preprint]{revtex4}

\usepackage{epsfig}
\usepackage{subfigure}
\usepackage{color}
\usepackage{amssymb}

\def\e{\begin{equation}}
\def\f{\end{equation}}

\begin{document}

\title{Electromagnetic cloaking of cylindrical objects by multilayer or uniform dielectric claddings}

\author{Constantinos A. Valagiannopoulos}
\author{Pekka Alitalo}
\affiliation{Department of Radio Science and Engineering / SMARAD Centre of Excellence, Aalto University, P.O. Box 13000, FI-00076 Aalto, Finland\\
{\rm E-mail: konstantinos.valagiannopoulos@aalto.fi, pekka.alitalo@aalto.fi}}

\date{\today}

\begin{abstract}
We show that dielectric or even perfectly conducting cylinders can be cloaked by a uniform or a layered dielectric cladding, without the need of any exotic or magnetic material parameters. In particular, we start by presenting a simple analytical concept that can accurately describe the cloaking effect obtained with conical silver plates in the visible spectrum. The modeled structure has been originally presented in [S. A. Tretyakov, P. Alitalo, O. Luukkonen, C. R. Simovski, Phys. Rev. Lett., vol.~103, p.~103905, 2009], where its operation as a cloak in the optical frequencies was studied only numerically. We model rigorously this configuration as a multi-layer dielectric cover surrounding the cloaked object, with excellent agreement to the simulation results of the actual device. The concept of using uniform or multilayer dielectric covers, with relative permittivities larger than unity, is then successfully extended to cloaking of impenetrable objects such as conducting cylinders.
\end{abstract}
\maketitle

\section{Introduction}
An electromagnetic cloak is a device that minimizes or even nullifies the effects of scattering from various objects by rendering them electromagnetically invisible to a detector. One well-known cloaking technique is the so-called ``transformation-optics'' method \cite{Leonhardt_Science, Pendry_Science, Nature_review, Jopt_review} which uses strongly inhomogeneous and highly anisotropic materials to perform complex transformations for the incident wave. Another cloaking technique is based on the cancellation of the scattering effects of a magnetodielectric object with a suitable metamaterial or plasmonic cladding \cite{Alu_transparency, Alu_review}, which neutralizes the polarization current of the primary scatterer.

Unfortunately, the actual construction of a cloaking device is very difficult with both the aforementioned approaches. Suggestions to overcome such a drawback includes, e.g., the use of transmission-line networks or other waveguiding structures instead of fictitious materials with exotic properties. \cite{Alitalo_review2} With transmission-line networks, the object to be electromagnetically hidden is strongly limited in size and geometry, \cite{TLcloak_Alitalo_APL2009, TLcloak_Alitalo_MOTL2009} but cloaking of impenetrable (e.g., perfectly metallic) objects has been shown to be possible with a set of conical conducting plates placed around a cylindrical cloaked region. \cite{VariousCloakingTechniques} This metal-plate cloaking has been practically realized in microwaves and numerically proven to be functional for a wide band of optical frequencies.\cite{MPcloak_Tretyakov_PRL2009} However, the underlying theory and the physical principles that govern the related phenomena have not been yet thoroughly understood.

In this work, we introduce a new, simple, nonmagnetic cloak. The idea originates from an analytical model of the previously reported metal-plate cloak operating in the visible range. \cite{MPcloak_Tretyakov_PRL2009} The used model is a simple mathematical concept comprised of multiple concentric cylinders of different dielectric permittivities. The choice of the permittivities has been made by considering the tapered conical lines, under the related plane-wave excitation, as a series connection of capacitors. The analytically evaluated response of the described model is shown to have a remarkable coincidence with the corresponding results obtained from full-wave simulations of the actual device for various values of input parameters.

Since the aforementioned simple model appears to resemble so successfully the real-world metal-plate cloaking structure, we tried to adopt an even simpler concept to cloak an impenetrable, perfectly electrically conducting (PEC) volume. This is a clearly more challenging case since one can hide practically anything in a PEC volume by avoiding to interact with the environment, which does not happen for penetrable objects. We introduce uniform or layered dielectric claddings which are optimized by sweeping the relative permittivity values of these covers. The scattering reduction achieved by the proposed devices is quite high, given their simple structure. Moreover, the demonstrated cloaking bandwidths can be considered as wide, a property attributed to the fact that the materials required in the claddings are simple dielectrics with relative permittivities larger than unity. It should be noted that most other cloaking methods for PEC objects, discussed in the literature, require magnetic material properties.

\section{Generic model of concentric cylinders}
\begin{figure}[ht]
\centering \epsfig{file=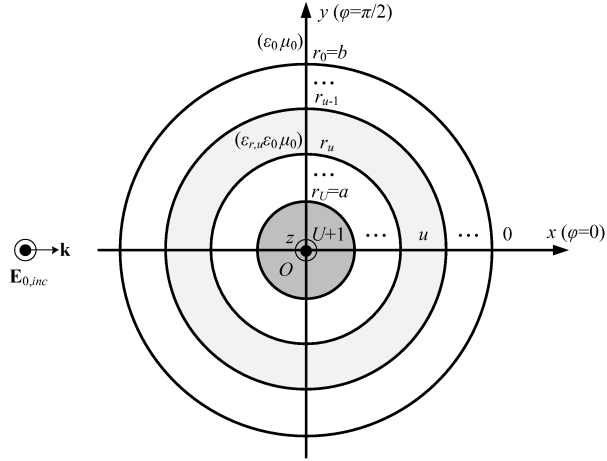, width=0.49\textwidth}
\caption{The geometry of the generic model comprised of several concentric cylindrical layers around the cloaked object. The cloaked object is situated in region $(U+1)$.}
\label{ModelGeometry}
\end{figure}

The proposed model to mimic the physical mechanism of wave propagation in metal-plate devices is fairly simple and its configuration is shown in Fig.~\ref{ModelGeometry}, where the used cylindrical coordinate system $(\rho, \phi, z)$ is also defined. There are just $U$ infinitely long cylindrical layers each of which is assigned to an integer number $u=1,\cdots,U$, occupying the region $r_{u-1}<\rho<r_u$ and being filled with magnetically inert dielectric materials of relative permittivities $\epsilon_{r,u}$. One can clearly note that $r_0=b$ and $r_U=a$, while the background medium is vacuum (with $\epsilon_0, \mu_0$ intrinsic electromagnetic parameters) and is taken as the region 0 ($\rho>b$) of our configuration with $\epsilon_{r,0}=1$. The internal cylinder, which corresponds to the region $(U+1)$ ($\rho<a$), can be either penetrable with relative dielectric constant $\epsilon_{r,(U+1)}$ or PEC. The assumed harmonic time dependence is of the form $e^{+j2\pi f t}$ ($f$ is the operating frequency) and is suppressed throughout the analysis.

We consider a plane wave excitation of electric field \cite{PlaneWaveExpansion}:
\begin{eqnarray}
{\bf E}_{0,inc}={\bf z}~ e^{-jk_0\rho\cos\phi}={\bf z} \sum_{n=-\infty}^{+\infty}j^{-n}J_n(k_0\rho)e^{jn\phi},
\label{eq:IncidentField}
\end{eqnarray}
where $k_0=2\pi f\sqrt{\epsilon_0\mu_0}$ is the free-space wavenumber and $J_n$ is the $n$-th ordered Bessel function. The total field in each layer $u=0,\cdots,(U+1)$ possesses only a $z$ component $E_{u}$ whose expression is given by:
\begin{eqnarray}
E_u=\sum_{n=-\infty}^{+\infty}\left[A_{n,u}J_n(k_u\rho)+B_{n,u}H_n(k_u\rho)\right]e^{jn\phi},
\label{eq:TotalFields}
\end{eqnarray}
where $k_u=k_0\sqrt{\epsilon_{r,u}}$, $H_n$ is the $n$-th ordered Hankel function of the second type and $A_{n,u}, B_{n,u}$ are sequences of unknown complex coefficients. After imposing the necessary boundary conditions at $\rho=r_u, u=0,\cdots, (U-1)$, one obtains the following relations connecting the coefficients of two adjacent layers:
\begin{eqnarray}
\left[\begin{array}{l} A_{n,u}\\B_{n,u} \end{array}\right]=
{\bf T}_{n,u}\cdot
\left[\begin{array}{l} A_{n,(u+1)}\\B_{n,(u+1)} \end{array}\right],
\label{eq:ElectromagneticBC}
\end{eqnarray}
for integer $n$. The explicit form of the transfer matrix \cite{TMatrices1, TMatrices2} ${\bf T}_{n,u}$ is shown in the two-column equation (\ref{eq:TransferMatrix}).

\begin{figure*}[t]
\hrulefill
\setcounter{equation}{3}
\begin{equation}
{\bf T}_{n,u}=\left[\begin{array}{cc} \frac{J'_n(k_{u+1}r_u)H_n(k_ur_u)k_{u+1}-H'_n(k_ur_u)J_n(k_{u+1}r_u)k_u}{J'_n(k_ur_u)H_n(k_ur_u)k_u-H'_n(k_ur_u)J_n(k_ur_u)k_u} &
\frac{H'_n(k_{u+1}r_u)H_n(k_ur_u)k_{u+1}-H'_n(k_ur_u)H_n(k_{u+1}r_u)k_u}{J'_n(k_ur_u)H_n(k_ur_u)k_u-H'_n(k_ur_u)J_n(k_ur_u)k_u} \\
\frac{J'_n(k_ur_u)J_n(k_{u+1}r_u)k_u-J'_n(k_{u+1}r_u)J_n(k_ur_u)k_{u+1}}{J'_n(k_ur_u)H_n(k_ur_u)k_u-H'_n(k_ur_u)J_n(k_ur_u)k_u} &
\frac{J'_n(k_ur_u)H_n(k_{u+1}r_u)k_u-H'_n(k_{u+1}r_u)J_n(k_ur_u)k_{u+1}}{J'_n(k_ur_u)H_n(k_ur_u)k_u-H'_n(k_ur_u)J_n(k_ur_u)k_u}
\end{array}\right].
\label{eq:TransferMatrix}
\end{equation}
\hrulefill
\end{figure*}

By successive application of (\ref{eq:ElectromagneticBC}) for $u=0,\cdots, (U-1)$, one arrives to:
\begin{eqnarray}
\left[\begin{array}{l} A_{n,0}\\B_{n,0} \end{array}\right]
={\bf T}_{n,0}\cdot{\bf T}_{n,1}\cdots{\bf T}_{n,(U-1)}\cdot\left[\begin{array}{l} A_{n,U}\\B_{n,U} \end{array}\right] \nonumber \\
=\left[\begin{array}{cc} M_{11}(n) & M_{12}(n)\\ M_{21}(n) & M_{22}(n) \end{array}\right]\cdot
\left[\begin{array}{l} A_{n,U}\\B_{n,U} \end{array}\right].
\label{eq:SuccessiveApplication}
\end{eqnarray}
Depending on what is the filling material of the core region $(U+1)$, namely dielectric (diel.) with permittivity $\epsilon_{r,(U+1)}$ or PEC, the following expression for the coefficients of the $U$ region (boundary conditions at $\rho=r_U=a$) is formulated:
\begin{eqnarray}
\left[\begin{array}{l} A_{n,U}\\B_{n,U} \end{array}\right]
=\left\{
\begin{array}{ll} {\bf T}_{n,U}\cdot\left[\begin{array}{c} A_{n,(U+1)}\\B_{n,(U+1)} \end{array}\right] & ,{\rm diel}. \\
                \left[\begin{array}{c} 1\\ -\frac{J_n(k_Ur_U)}{H_n(k_Ur_U)} \end{array}\right]A_{n,U}  & ,{\rm PEC}
\end{array}.
\right.
\label{eq:InternalBoundary}
\end{eqnarray}
By inspection of (\ref{eq:IncidentField}), it is directly obtained that $A_{n,0}=j^{-n}$, while the physical demand for bounded field into the cloaked region, is translated into: $B_{n,(U+1)}=0$. Therefore, the scattering field by the device into the vacuum background area can be readily derived through the related coefficient $B_{n,0}$ by combining (\ref{eq:SuccessiveApplication}), (\ref{eq:InternalBoundary}) in each case:
\begin{eqnarray}
B_{n,0}
=\left\{
\begin{array}{ll}  j^{-n}\frac{M_{21}(n)\left[{\bf T}_{n,U}\right]_{11}+M_{22}(n)\left[{\bf T}_{n,U}\right]_{21}}
                              {M_{11}(n)\left[{\bf T}_{n,U}\right]_{11}+M_{12}(n)\left[{\bf T}_{n,U}\right]_{21}}
                              & , {\rm diel}. \\
                   j^{-n}\frac{M_{21}(n)-M_{22}(n)\frac{J_n(k_Ur_U)}{H_n(k_Ur_U)}}
                              {M_{11}(n)-M_{12}(n)\frac{J_n(k_Ur_U)}{H_n(k_Ur_U)}}
                              & , {\rm PEC}
\end{array}.
\right.
\label{eq:ScatteringCoefficient}
\end{eqnarray}
The notation $\left[{\bf D}\right]_{vw}$ is used for the $(v,w)$ element of the matrix ${\bf D}$.

\begin{figure}[ht]
\centering \epsfig{file=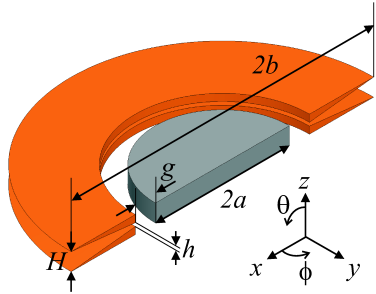, width=0.3\textwidth}
\caption{(Color online) The physical configuration of a single conical metal-plate cell. One half (cut along the $xz$-plane) of a single cell is shown.}
\label{ConicalComponent}
\end{figure}

\section{Optical cloak made of conical plates}
In this section, we are going to test the model analyzed above on how accurately does it describe the operation of an actual device. It has been shown that the so-called metal-plate cloak configuration, already constructed for radio frequencies, can work in the optical band as well. \cite{MPcloak_Tretyakov_PRL2009} The optical device is comprised of periodically stacked, solid, conical, silver plates (of outer diameter 2$b$) positioned around the cylindrical cloaked region (of diameter 2$a$) with a small air gap (of thickness $g$) in between. One period of this structure is shown in Fig.~\ref{ConicalComponent} and it is a (hollow) waveguide, with linearly decreasing height from $H$ (at $\rho=b$) to $h$ (at $\rho=a$), leading the $z$-polarized fields around the cloaked object. In a specific case \cite{MPcloak_Tretyakov_PRL2009}, the constructing material of the plates is silver, the (optimal) operating frequency $f_0\cong 590$ THz, while the physical dimensions are given by: $b=113$ nm, $a=50$ nm, $g=15$ nm, $H=13$ nm and $h=2.5$ nm. It should be stressed that the cloaked region is filled with silver too, while the frequency-dependent permittivity of this substance $\epsilon_{r,silver}=\epsilon_{r,silver}(f)$ is well known. \cite{SilverPermittivity} Note that the permittivity of silver is negative close to $f=f_0$, namely: $\Re[\epsilon_{r,silver}(f_0)]\cong-10$.

\begin{figure}[ht]
\centering \epsfig{file=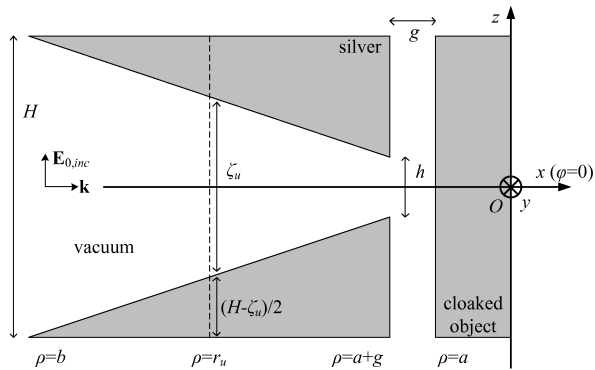, width=0.49\textwidth}
\caption{The side view of a single conical metal-plate cell.}
\label{SideView}
\end{figure}

With reference to the model analyzed above, we make use of the formulas corresponding to a dielectric core since the region $(U+1)$ is solid silver; therefore, $\epsilon_{r,(U+1)}=\epsilon_{r,silver}$. In addition, the $U$-th layer is the air gap; thus, $\epsilon_{r,U}=1$ and $r_{U-1}=a+g$. It is apparent that our model does not take into account the periodic dimension variation of the actual configuration with respect to $z$, since the stack of conical plates is replaced by a structure of homogeneous concentric cylinders. However, the radial inhomogeneity (contrary to the axial one) of the cloaking device is possible to be imitated by choosing properly the dielectric permittivities of the cylindrical layers. But how can one compute $\epsilon_{r,u}$ for $u=1,\cdots,(U-1)$ in a correct and reliable way?

In Fig.~\ref{SideView}, one observes the side view of half the biconical cell as the shape possesses cylindrical symmetry. Assume that the cylindrical layer in the vicinity of the representative surface $\rho=r_u$ (containing both the corresponding vacuum aperture and the related solid silver plates), would be replaced by a (locally) homogeneous dielectric, with the same external dimensions, of relative permittivity $\epsilon_{r,u}$, whose value should be estimated. From the similarity of triangles, the height $\zeta_u$ of the representative vacuum aperture is found equal to:
\begin{eqnarray}
\zeta_u=H\frac{r_u-a-g}{b-a-g}+h\frac{b-r_u}{b-a-g}.
\label{eq:RepresentativeHeight}
\end{eqnarray}
If the inclination of the tapered plates is relatively low (which is the case for the specific choice of $b, a, g, H,h$), then the $z$-polarized electric field will be almost normal to the sloping boundaries separating the silver  from the vacuum. As a result, the arbitrary cross section at $\rho=r_u$ can be considered as a series connection of three capacitors: two with the height $(H-\zeta_u)/2$ filled with silver and one (placed in between) with height $\zeta_u$ filled with vacuum. Accordingly, this serial cluster can be replaced by a single capacitor with a dielectric material of relative permittivity $\epsilon_{r,u}$, based on the well-known equivalent capacitance formula: $\frac{H}{\epsilon_{r,u}}=\frac{(H-\zeta_u)/2}{\epsilon_{r,silver}}+\frac{\zeta_u}{1}+\frac{(H-\zeta_u)/2}{\epsilon_{r,silver}}$. In this sense, the rigorous expression for $\epsilon_{r,u}$ is hyperbolic with respect to both $r_u$, $\zeta_u$ and is given as follows:
\begin{eqnarray}
\epsilon_{r,u}=\frac{H\epsilon_{r,silver}}{H+(\epsilon_{r,silver}-1)\zeta_u}.
\label{eq:EquivalentPermittivity}
\end{eqnarray}
We use layers of the same thickness (due to the constant inclination of the silver conical plates), namely: $r_u=b-\frac{u}{U-1}(b-a-g)$ for $u=1,\cdots,(U-1)$. After having defined all the parameters of the theoretical structure ($\epsilon_{r,u}, r_u$) based on the actual quantities of the real configuration ($b, a, g, H, h, \epsilon_{r,silver}$), we can develop the corresponding model and quantify its efficiency. It is remarkable that (for the given ranges of the input parameters) all the relative permittivities $\epsilon_{r,u}$ are positive and greater than unity.

\begin{figure}[ht]
\centering \epsfig{file=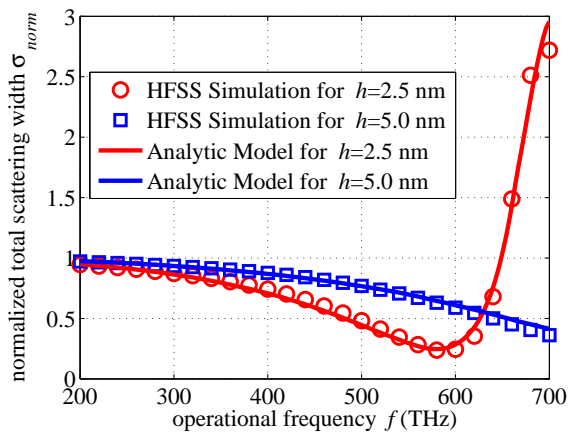, width=0.49\textwidth}
\caption{(Color online) The total scattering width of the metal-plate structure normalized by the corresponding quantity of the uncloaked object as function of the operating frequency. The HFSS simulations for the actual device are compared with the results obtained through the analytical model implementation. Plot parameters: $b=113$ nm, $a=50$ nm, $g=15$ nm, $H=13$ nm, $U=8$.}
\label{FigA}
\end{figure}

A crucial quantity for the operation and the performance of a cloaking device is the total scattering width of the whole cylindrical structure normalized by the corresponding width of the uncloaked one. The smaller is the magnitude, the better the device serves its purpose. The quantity is defined by:
\begin{eqnarray}
\sigma_{norm}=\frac{\int_0^{2\pi}\left|\sum_{n=-N}^N
B_{n,0}j^ne^{jn\phi}\right|^2
d\phi}{\int_0^{2\pi}\left|\sum_{n=-N}^N
B'_{n,0}j^ne^{jn\phi}\right|^2 d\phi},
\label{eq:TotalScatteringWidth}
\end{eqnarray}
where $N$ is a sufficiently large integer to achieve convergence for the series and $B'_{n,0}$ denote the coefficients of the scattering field for the uncloaked cylinder. In Fig.~\ref{FigA} we show the variation of $\sigma_{norm}$ with respect to the operating frequency for two alternative heights $h=2.5$ and $h=5$ nm. In each case, we perform a numerical simulation of the actual metal-plate device with ANSYS HFSS full-wave software. \cite{HFSS} The obtained numerical results are compared to those derived through implementation of the corresponding multi-layered analytical model, in the way described above. There is an excellent agreement between the two sets of data despite the fact that they describe two completely different structures (simulation of an axially inhomogeneous real device and rigorous solution to an axially homogeneous multilayered configuration). This remarkable coincidence between so dissimilar configurations demonstrates the success of the adopted model. When it comes to the results themselves, there are large frequency bands where the magnitude $\sigma_{norm}$ takes values much smaller than unity. In addition, the metal-plate cloaking for $h=5$ nm is functional at larger frequencies than the device with $h=2.5$ nm does.

\begin{figure}[ht]
\centering \epsfig{file=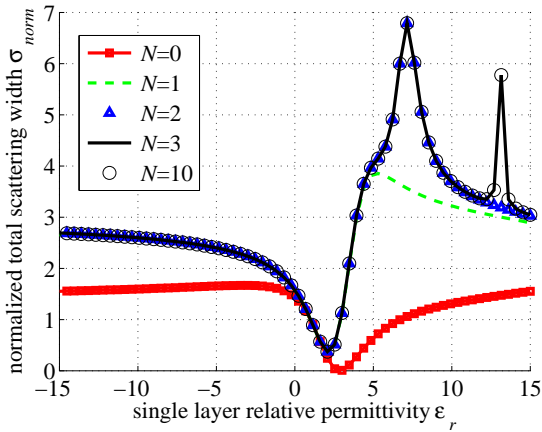, width=0.49\textwidth}
\caption{(Color online) The normalized total scattering width of a silver rod surrounded by a homogeneous cladding as function of the relative permittivity of the cladding. Plot parameters: $b=113$ nm, $a=50$ nm, $f_0=590$ THz, $U=1$.}
\label{FigG}
\end{figure}

In the considered numerical example, the electrical dimension of the structure is relatively small which could make one think that just the omnidirectional ($N=0$) term of the sums in (\ref{eq:TotalScatteringWidth}), is sufficient to evaluate the quantity $\sigma_{norm}$. If such an argument was correct, then the cloaking behavior demonstrated above would be attributed to the well-known ``scattering cancellation'' phenomenon. \cite{Alu_transparency, Alu_review} We believe that this is not the case since the electrical size of the structure is not so small. To support this statement, we assume the corresponding ``scattering cancellation'' device for the silver cloaked object of radius $a$. This is a homogeneous cladding of relative permittivity $\epsilon_r$ and radius $b$ without an air gap. In Fig.~\ref{FigG}, we evaluate the normalized total scattering width $\sigma_{norm}$ from (\ref{eq:TotalScatteringWidth}) as function of $\epsilon_r$ for several truncation limits $N$. We are mainly interested in the interval $0<\epsilon_r<5$ where smaller $\sigma_{norm}$ are observed and it is clear that for $N\ge 2$, the result does not change significantly. However, the first harmonic ($N=1$) is necessary to be included in the computation since the zeroth term is minimized (and more specifically nullified) at larger $\epsilon_r$. The nullification of the omnidirectional term is the ``scattering cancellation'' solution and it is different from the observed one.

\section{Cloaking of conducting cylinders}
As was shown in the previous section, a multilayer as well as a single layer dielectric cover can be used to reduce the scattering from a cylinder composed of an $\epsilon_r$--negative material. The same can be of course achieved by using the well-known ``scattering cancellation method''. However, this technique has not been used so far to hide a conducting object by a layered or a uniform cover made of dielectrics with $\epsilon_r>1$. Here we show, by using the analytical model described in Section~II, that such a case is possible at least for conducting cylinders of moderate electrical size.

The electromagnetic fields weakly penetrate the cloaked silver cylinder studied in the previous section, so it is expected that also conducting cylinders, i.e., impenetrable objects, could also be partially cloaked with simple multilayer or even uniform covers. To present a generalized case, in the following we normalize the dimensions and the results to the free-space wavelength $\lambda_0$. We study an electrically thin, PEC cylinder with radius $a=\lambda_0/10$ (the electrical size of this cylinder is of the same order as the cloaked object in the previous section). We assign a dielectric cover around this cylinder and consider three cases: (i) a uniform cover with constant $\epsilon_r$, (ii) a multilayer cover with linearly varying $\epsilon_{r,u}$, (iii) a multilayer cover with hyperbolically varying $\epsilon_{r,u}$ (as in (\ref{eq:EquivalentPermittivity})). For simplicity, we choose $b=2a$ in all the studied cases. However, the choice of $b$ and $a$ can be made freely; the resulting value of (optimal) $\epsilon_{r,U}$ of the cover material simply changes if the values of $a$ and $b$ are changed.

\begin{figure}[t!]
\centering \epsfig{file=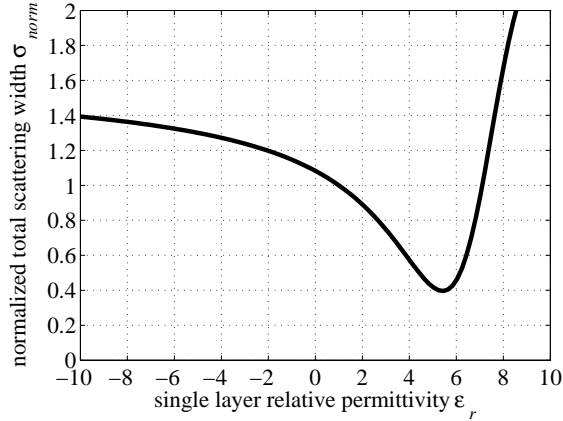, width=0.49\textwidth}
\caption{Sweep of $\epsilon_r$ to find the optimal value for cloaking at the frequency $f_0$ with fixed values of $a$ and $b$. The uniform ($U=1$) cloak has a thickness $(b-a)$ (with $b=2a=\lambda_0/5$) and is made of a dielectric material with $\epsilon_r$.}
\label{epsr_opt1}
\end{figure}

In the first case, we assign a constant $\epsilon_r$ to the material surrounding the PEC cylinder. The analytical model of Section~II is used to optimize the value of $\epsilon_r$ (see Fig.~\ref{epsr_opt1}). It is evident that for these values of $a$ and $b$, the optimal value is $\epsilon_r=5.42$ and with that, the normalized total scattering width of the cloaked PEC cylinder is slightly less than 0.4, i.e., the cloak reduces the total scattering width $\sigma_{norm}$ of the PEC cylinder by more than 60~\%.

\begin{figure}[t!]
\centering \epsfig{file=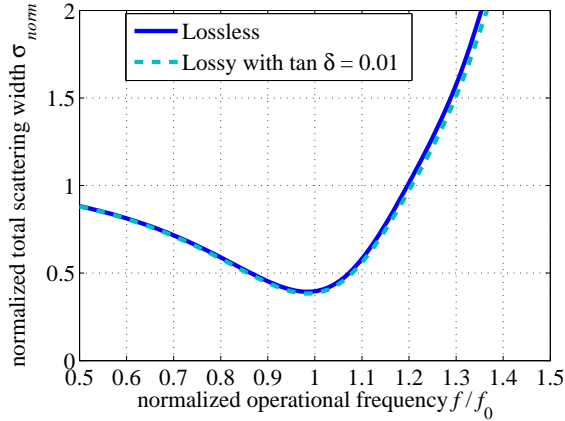, width=0.49\textwidth}
\caption{(Color online) Normalized total scattering width as a function of the normalized frequency. The uniform ($U=1$) cloak has a thickness $(b-a)$ (with $b=2a=\lambda_0/5$) and is made of a dielectric material with $\epsilon_r=5.42$.}
\label{constant_epsr}
\end{figure}

The frequency dependence of the normalized total scattering width is illustrated in Fig.~\ref{constant_epsr}, demonstrating a reasonably broadband cloaking effect: the relative bandwidth where the total scattering width of the PEC cylinder is reduced to the half or less, is about 21~\%. It is quite interesting to note that moderate losses do not deteriorate the cloaking effect; with a loss tangent of 0.01, the cloaking effect is actually slightly improved compared to the lossless case.

\begin{figure}[t!]
\centering \epsfig{file=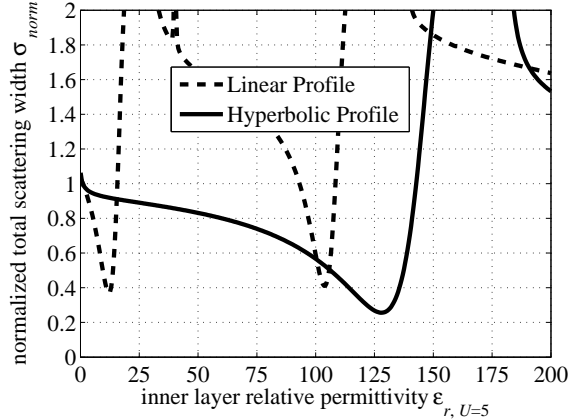, width=0.49\textwidth}
\caption{Sweep of $\epsilon_{r,U=5}$ to find the optimal value for cloaking at the frequency $f_0$ with fixed values of $a$ and $b=2a=\lambda_0/5$. Linear and hyperbolic profiles $(U=5)$ are considered.}
\label{epsr_opt2}
\end{figure}

In the second scenario, we model the case of Fig.~\ref{ModelGeometry} for $U=5$ and the value of $\epsilon_{r,u}$ depending on the layer number $u$ linearly. To find the optimal values of $\epsilon_{r,u}$ for cloaking at $f=f_0$, we plot $\sigma_{norm}$ as a function of $\epsilon_{r,5}$, i.e., the maximum value of $\epsilon_r$, while $\epsilon_{r,u}$ (for $u=1,2,3,4$) is linearly decreasing from $\epsilon_{r,5}$ to $\epsilon_{r,0}=1$. The result shown in Fig.~\ref{epsr_opt2} demonstrates that for the chosen dimensions $a$ and $b$, the lowest positive value of $\epsilon_{r,5}$, corresponding to a minimum in the normalized total scattering width, is $\epsilon_{r,5}=12.1$. In that case, the relative permittivities of the five-layer cloak are as shown in Table~I.

\begin{table}[t!]
\centering \caption{Values of relative permittivities with linearly changing $\epsilon_r$.}
\label{table2}
\begin{tabular}{|c|c|c|c|c|c|} \hline
$u$ & 1  &  2  &  3  & 4 & 5\\
\hline
$\epsilon_{r,u}$ & 3.22 & 5.44 & 7.66 & 9.88 & 12.1 \\
\hline
\end{tabular}
\end{table}

\begin{figure}[t!]
\centering \epsfig{file=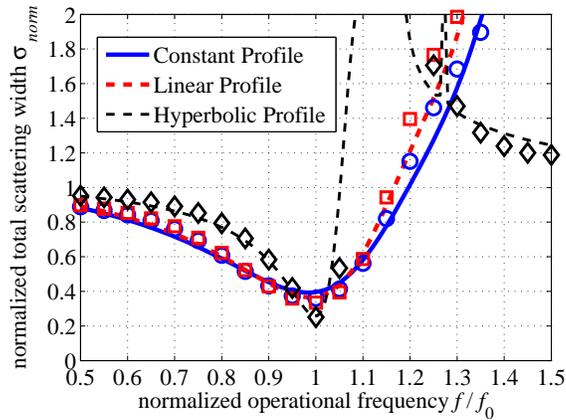, width=0.49\textwidth}
\caption{(Color online) Normalized total scattering width as a function of the normalized frequency for constant, linear, and hyperbolic profile $(U=5)$. The circles, squares, and diamonds denote the numerical results of each of the previous cases, respectively.}
\label{varying_epsr}
\end{figure}

With the values of Table~I, the normalized total scattering width as function of the frequency looks as shown in Fig.~\ref{varying_epsr}. In the same graph, the curve for the constant-permittivity cladding is depicted for comparison and also the corresponding numerical results obtained with ANSYS HFSS (showing good agreement with our analytical findings) have been attached.

Finally, we study the third scenario, i.e., the case with $\epsilon_r$ depending hyperbolically on the layer number $u=0,\cdots,U$, where again $U=5$. The variation of the normalized total scattering width $\sigma_{norm}$ with respect to $\epsilon_{r,5}$, is shown in Fig.~\ref{epsr_opt2}, resulting in the value $\epsilon_{r,5}=128$ for optimal operation at $f_0$. With this value, the relative permittivities of the five-layer cloak are as shown in Table~II. The frequency dependence of the normalized total scattering width is illustrated in Fig.~\ref{varying_epsr}. Again, we verify our analytical evaluations by plotting the simulation results in the same figure.

\begin{table}[t!]
\centering \caption{Values of relative permittivities with hyperbolically changing $\epsilon_r$.}
\label{table3}
\begin{tabular}{|c|c|c|c|c|c|} \hline
$u$ & 1  &  2  &  3  & 4 & 5\\
\hline
$\epsilon_{r,u}$ & 1.25 & 1.66 & 2.47 & 4.85 & 128 \\
\hline
\end{tabular}
\end{table}

Comparing the three curves of Fig.~\ref{varying_epsr}, we can conclude that changing the profile of $\epsilon_{r,u}$ from constant to linear and hyperbolic, improves the obtained scattering reduction at the frequency $f_0$, but at the same time, the bandwidth of efficient cloaking decreases. Concerning practical issues, the constant and linear profiles are easily realizable (the required values of $\epsilon_{r,u}$ are feasible), whereas the hyperbolic profile is far from practical. However, the cloaking performance with the constant profile is not much different from the linear profile, so it may not be worth the increased complexity to use even the linear one.

It is clear that cloaking of PEC objects with simple dielectric covers is far from ideal cloaking such that is in theory possible with, e.g., ``transformation-optics''. However, it is evident that the cloaking efficiencies presented in this work, are comparable to the experimental and numerical results obtained with various other cloaking techniques that have been realized with composite metamaterials. \cite{alu_exp, smith_exp}

\section{Conclusions}
We have presented a very simple analytical concept, based on transfer matrices at multilayered cylindrical structures. It has been found that this model describes accurately the previously reported cloaking effect obtained with conical silver plates in the visible frequency range. The effectiveness of the analytical model is verified by comparing the results of the normalized total scattering widths originating from it, to results obtained by numerical simulation software. The fidelity of the proposed concept allows it to be used in device design and to save computational resources due to its simplicity. The same analytical model is also used to demonstrate that, surprisingly, cloaking of impenetrable (perfectly conducting) objects is possible with simple dielectric covers whose relative permittivity surpasses unity. Such a property renders this type of electromagnetic configurations easily realizable, unlike most other cloaking devices reported in the literature.

\section{Acknowledgments}

The authors wish to thank Prof. S. Tretyakov for useful advice and discussions related to the topic of this paper. The work of P. Alitalo was supported by the Academy of Finland via postdoctoral project funding.


\begin{thebibliography}{99}

\bibitem{Leonhardt_Science}
U.~Leonhardt,
\textit{Science} \textbf{312}, 1777 (2006).

\bibitem{Pendry_Science}
J.~B. Pendry, D. Schurig, and D.~R.~Smith,
\textit{Science} \textbf{312}, 1780 (2006).

\bibitem{Nature_review}
H.~Chen, C.~T.~Chan, and P.~Sheng,
\textit{Nature Materials} \textbf{9}, 387 (2010).

\bibitem{Jopt_review}
S.~Guenneau, R.~C.~McPhedran, S.~Enoch, A.~B.~Movchan, M.~Farhat, and N.-A.~P.~Nicorovici,
\textit{J. Opt.} \textbf{13}, 024014 (2011).

\bibitem{Alu_transparency}
A.~Al$\rm{\grave{u}}$ and N.~Engheta,
\textit{Phys. Rev. E} \textbf{72}, 016623 (2005).

\bibitem{Alu_review}
A.~Al$\rm{\grave{u}}$ and N.~Engheta,
\textit{J.~Opt.~A} \textbf{10}, 093002 (2008).

\bibitem{Alitalo_review2}
P.~Alitalo and S.~Tretyakov,
\textit{Proc. IEEE} \textbf{99}, 1646 (2011).

\bibitem{TLcloak_Alitalo_APL2009}
P.~Alitalo, F.~Bongard, J.-F.~Z\"urcher, J.~Mosig, and S.~Tretyakov,
\textit{Appl. Phys. Lett.} \textbf{94}, 014103 (2009).

\bibitem{TLcloak_Alitalo_MOTL2009}
P.~Alitalo, O.~Luukkonen, J.~R.~Mosig, and S.~A.~Tretyakov,
\textit{Microw. Opt. Technol. Lett.} \textbf{51}, 1627 (2009).

\bibitem{VariousCloakingTechniques}
P.~Alitalo, H.~Kettunen, and S.~A.~Tretyakov,
\textit{J. Appl. Phys.} \textbf{107},  03490 (2010).

\bibitem{MPcloak_Tretyakov_PRL2009}
S. Tretyakov, P. Alitalo, O. Luukkonen, and C. Simovski,
\textit{Phys. Rev. Lett.} \textbf{103}, 103905 (2009).

\bibitem{PlaneWaveExpansion}
C.~A.~Valagiannopoulos,
\textit{Int. J. Ant. Propagat.} \textbf{2009}, 301461 (2009).

\bibitem{TMatrices1}
C.~A.~Valagiannopoulos and N.~L.~Tsitsas,
\textit{J. Opt. Soc. Am. A} \textbf{26}, 870 (2009).

\bibitem{TMatrices2}
C.~A.~Valagiannopoulos and N.~L.~Tsitsas,
\textit{Electromagnetics} \textbf{28}, 243 (2008).

\bibitem{SilverPermittivity}
P.~B.~Johnson and R.~W.~Christy,
\textit{Phys. Rev. B} \textbf{6}, 4370 (1972).

\bibitem{HFSS}
Homepage of Ansys HFSS on the internet: www.ansoft.com/products/hf/hfss.

\bibitem{alu_exp}
B.~Edwards, A.~Al$\rm{\grave{u}}$, M.~Silveirinha, and N.~Engheta,
\textit{Phys. Rev. Lett.} \textbf{103}, 153901 (2009).

\bibitem{smith_exp}
N.~Kundtz, D.~Gaultney, and D.~R.~Smith,
\textit{New J. Phys.} \textbf{12}, 043039 (2010).

\end{thebibliography}
\end{document}